\documentclass[12pt]{article}
\usepackage[utf8]{inputenc}
\usepackage{enumerate}
\usepackage{xcolor}
\usepackage{amsmath}
\usepackage{amsfonts}
\usepackage{amssymb}
\usepackage{graphicx}
\usepackage{authblk}
\usepackage{graphics, setspace}
\newcommand{\mathsym}[1]{{}}
\newcommand{\unicode}[1]{{}}

\newcommand{\bea}{\begin{eqnarray*}}
\newcommand{\eea}{\end{eqnarray*}}
\newcommand{\be}{\begin{equation}}
\newcommand{\ee}{\end{equation}}

\begin{document}

\providecommand{\abs}[1]{\lvert#1\rvert}

\title{Energy preserving boundary conditions in field theory}
\author{Manuel Asorey and Fernando Ezquerro}
\affil{ Centro de Astropart\'{\i}culas y F\'{\i}sica de Altas Energ\'{\i}as, Departamento de F\'{\i}sica Te\'orica.
Universidad de Zaragoza,  E-50009 Zaragoza, Spain}

\date{}
\maketitle

\begin{abstract}
 The dynamics of classical field theories is usually governed by  field equations, but when fields
 are constrained to bounded domains it is also dependent on its boundary conditions.
Usually boundary conditions are constrained  by the requirement of preserving the maximal symmetry
of the system. In the case of charged particles the symmetry is $U(1)$, but there are many fields
 (e,g, electromagnetic fields) which are neutral and charge conservation does not constraint its boundary 
 conditions. 
In this paper we explore the most general boundary conditions that preserve another symmetry that all
relativistic field theories do preserve: space-time translations. In particular the families of boundary conditions
 of isolated systems which  preserve energy for
scalar, electromagnetic and Yang-Mills field theories. We point out the global  properties of the space of
all possible boundary conditions of  confined systems in two special domains. 
We also explore the  connection between  energy  preserving  and charge preserving boundary conditions.
\end{abstract}

\section{Introduction}

The new role of quantum boundary effects 
is boosting in condensed matter  a new era of quantum technologies.
Indeed, the presence of plasmons and other surface effects
in metals and dielectrics  \cite{ZZX}, the appearance of edge currents in the Hall effect \cite{laughlin}
--\cite{Hat}, and the discovery of new edge effects  in topological insulators \cite{KM}
--\cite{BernHugh}  and Weyl semiconductors \cite{Hassan}
 have  very rich potential implications. 

Although boundary effects arise in quantum physics since the early days of the theory, the role of 
boundary effects in quantum field theory have a  much later development.

  Boundary effects  also
appear today as an essential ingredient in fundamental physics. Since
the discovery of the Casimir effect \cite{casimir}  they also arise behind  new quantum effects like
 Hawking radiation,  Black Hole horizons effects,
topological defects, topology change \cite{bala}
--\cite{survey}  and  holography in the AdS/CFT correspondence.

The increasing relevant role of boundary effects is demanding a comprehensive theory of
boundary conditions.
In spite of the fact that quite a lot of work has been devoted to establish the foundations
of the quantum theory, a comprehensive theory of boundary conditions for quantum field theories
is still missing. 
This gap was filled 
 by using the unitarity principle for time evolution  \cite{aim}
that was further extended to quantum field theories \cite{agamc,amc} . 
This first global analysis was based on the preservation
of unitarity for time evolution. 
The generalization for relativistic field theories requires a change of the  basic principles, from  unitarity to
the preservation of the $U(1)$ symmetry which is responsible of
electric charge conservation  \cite{Bangalore}. However, this principle does not apply to neutral or gauge fields where the new approach does not provide any fundamental law to be preserved. 

In this paper we address the analysis of  the theory of boundary conditions in field theories confined  in isolated domains
 based  only on the requirement of conservation of energy.  This method applies to any  bosonic or fermionic relativistic field theory including neutral fields like gauge fields, with the only exception  of gravitation and topological field theories.

In section 2 we start our analysis with the neutral scalar fields in the case of one boundary wall. In section 3 we study the case of charged scalar fields. In section 4 we complete the discussion about charged scalar fields showing the compatibility between the condition of charge conservation and energy conservation. In section 5 we generalize this approach to Yang-Mills theories developing the theory of boundary conditions which preserve energy and gauge invariance.
In section 6 we address the case of  interacting theories of matter and gauge fields.
In section 7 we study the case of two different electromagnetic active media separated by a planar surface. Finally, in section 8 we summarize the  applications and conclusions of this work.

\section{Neutral scalar fields in half space}
The dynamics of  a real scalar field is governed by the Lagrangian density
\begin{equation}
	\mathcal{L}=\frac{1}{2}\partial_\mu\phi\partial^\mu\phi - \frac{1}{2} m^2 \phi^2 - V(\phi),
\end{equation}
where $V(\phi)$ is any arbitrary local potential function.
The space-time translation symmetry induces by Noether theorem four conservation laws
\begin{equation}
\partial^\mu	T_{\mu\nu}= 0,
\end{equation}
where the energy-momentum tensor is given by
\begin{equation}
	T_{\mu\nu}=\frac{1}{2}\partial_\mu\phi\partial_\nu\phi-\frac{1}{2}\eta_{\mu \nu} \partial_\alpha\phi\partial^\alpha\phi  + \frac{1}{2} m^2 \phi^2\eta_{\mu \nu} + V(\phi) \eta_{\mu \nu},
\end{equation}
 $\eta_{\mu \nu}$being the Minkowski metric.
 In particular when $\nu=0$ we get the energy  conservation law 
 \begin{equation}\nonumber
\partial^\mu	T_{\mu 0}= \partial_t {\mathcal{E}} + \partial ^i P_i =0,
\end{equation}
where 
\begin{equation}\nonumber
{\mathcal E}=  \frac{1}{2}\partial_t \phi \partial_t \phi+ \frac{1}{2}\partial_i\phi\partial_i\phi  + \frac{1}{2} m^2 \phi^2 + V(\phi)
\end{equation}
is the energy density and 
\begin{equation}\nonumber
P_i=\frac{1}{2}\partial_0  \phi\, \partial_i \phi ,
\end{equation}
 the momentum density of the field (we assume that  $c=1$ from now on).
Thus, for any bounded domain $\Omega$ with regular boundary $\partial \Omega$ the variation of the intrinsic field energy is
\begin{equation}\nonumber
\frac{d}{d t}{E}_\Omega = \int_\Omega \partial_t {\mathcal E}\ d^3x=- \int_\Omega  \partial^i P_i\ d^3x =-\int_{\partial \Omega} n^i  P_i\ d\sigma_{_{\partial \Omega}},
\end{equation}
where ${\mathbf n}= (n^i)$ denotes the normal vector to the boundary surface $\partial \Omega$.

Let us consider a simple case where $\Omega$ is just a half-space $\Omega=\{ {\mathbf x}=(x^1,x^2, x^3) | x^3\geq  0   \} $ whose boundary $\partial \Omega=\{ {\mathbf{x}}=(x^1,x^2, 0)   \}$ is the plane perpendicular to the vector ${\mathbf{n}} =(0,0,-1)\in {\mathbb{R}}^3$. In that case the conservation of energy implies that
\begin{equation} \nonumber
	\frac{d}{d t}{E}_\Omega =\iint T_{03}\ dx^1 dx^2=\iint \dot\varphi\,\varphi'\ dx^1 dx^2=0,
\end{equation}
where $$ \dot \varphi = \partial_t \phi\Big|_{\partial\Omega} $$ and $$\varphi' =\partial_3 \phi\Big|_{\partial\Omega}$$ are the boundary values of  the time derivative and  the  normal derivative  of the fields $\phi$ across the boundary  $\partial\Omega$\footnote{In dimensions higher than 2 there are some technical difficulties concerning the regularity of boundary values \cite{amc} .  In this paper we will restrict ourselves to cases of  regular boundary conditions} . 
If we consider only homogeneous boundary conditions which are invariant under translations along the boundary  plane $\partial\Omega$, the conservation of energy requires that 
\begin{equation}
\dot\varphi\, \varphi'=0
\end{equation}
which has two solutions $\varphi'=0$ and $\dot\varphi=0$.  The first solution corresponds to Neumann boundary condition which in string theory is the usual boundary condition for open strings, whereas the second solution includes the Dirichlet  boundary conditions which corresponds to D-branes in that theory \cite{Polchinski}.

\section{Charged scalar fields}
Let us now consider the case of   complex scalar fields with Lagrangian density
\begin{equation}
	\mathcal{L}=\frac{1}{2}\partial_\mu\phi^\ast \partial^\mu\phi - \frac{1}{2} m^2 |\phi|^2 - V(|\phi|),
\end{equation}
where $V(\phi)$ is any arbitrary local density potential function.
Using the same arguments as in the case of real scalar fields we get that energy conservation requires the vanishing of 
\begin{equation}\label{energycon}
	\dot{\varphi^*}\varphi'+\varphi^{*\prime}\dot\varphi=0.
\end{equation}
After the change of variables
\begin{equation}\nonumber
	\psi_1=\begin{pmatrix}
		\varphi'+\dot\varphi\\
		\varphi^{*\prime}+\dot{\varphi^*}
	\end{pmatrix}\hspace{1cm}\psi_2=\begin{pmatrix}
	\varphi'-\dot\varphi\\
	\varphi^{*\prime}-\dot{\varphi^*}
\end{pmatrix},
\end{equation}
the vanishing condition becomes
\begin{equation} \nonumber
\abs{\psi_1}^2-\abs{\psi_2}^2
=4\left(\varphi^{*\prime}\dot\varphi+\dot{\varphi^*}\varphi'\right).
\end{equation}
If we restrict ourselves to boundary conditions which are  translation invariant along the boundary the most general solution   satisfies 
\begin{equation} \label{bce}
\left|\begin{pmatrix}
	\varphi'+\dot\varphi\\
	\varphi^{*\prime}+\dot{\varphi^*}
\end{pmatrix}\right|^2=\left| \begin{pmatrix}
\varphi'-\dot\varphi\\
\varphi^{*\prime}-\dot{\varphi^*}
\end{pmatrix}\right|^2
\end{equation}
and is given by
\begin{equation} \label{bce1}
	\begin{pmatrix}
		\varphi'+\dot\varphi\\
		\varphi^{*\prime}+\dot{\varphi^*}
	\end{pmatrix}=U \begin{pmatrix}
		\varphi'-\dot\varphi\\
		\varphi^{*\prime}-\dot{\varphi^*}
	\end{pmatrix},
\end{equation}
where $U$ is an arbitrary  2x2 unitary matrix.
Conjugation of equation  (\ref{bce1}) leads to
\begin{equation}\nonumber
		\begin{pmatrix}
		\varphi^{*\prime}+\dot{\varphi^*}\\
		\varphi'+\dot\varphi
	\end{pmatrix}=U^* \begin{pmatrix}
		\varphi^{*\prime}-\dot{\varphi^*}\\
		\varphi'-\dot\varphi
	\end{pmatrix}; \quad \sigma_1 \psi_1=U^*\sigma_1\psi_2, \quad \psi_1=\sigma_1U^*\sigma_1\psi_2.
\end{equation}
This implies that  the unitary matrix $U$ has to satisfy an extra  condition
\begin{equation}\label{extra2}
U=\sigma_1U^*\sigma_1.
\end{equation}

The meaning of this restriction is that the unitary matrix $U$   has to belong  also to the $O(1,1)$ rotation group, because 
from (\ref{extra2}) it follows that  $U$ leaves the $(1,1)$ metric 
\begin{equation} \nonumber
\sigma_1=
\begin{pmatrix}
		0&1\\
		1&0 
	\end{pmatrix}
\end{equation}
invariant. i.e.
\begin{equation} \nonumber
U^\perp \sigma_1 U= \sigma_1.
\end{equation} 

Thus, the set of  homogeneous local boundary conditions  is   given by the one-dimensional subgroup 
of  unitary matrices $G=O(1,1)\cap U(2)$\footnote{This is one of three maximal compact subgroups of $U(2)$\cite{afg} }  which has two disjoint components 
$G=G_+\cup G_-$ whose matrices only differ  by the sign of their determinant $ \det U=\pm 1$. The   component $G_+$
contains all the matrices of the form
\begin{equation}\label{one} 
	U_+(a)= e^{ \, i a \sigma_3}\qquad  a\in [0, 2\pi), 
\end{equation}
that are continuously connected with the identity,
whereas the other component $G_-$ is given by all the matrices
\begin{equation}\label{two}
	U_-(b)=\sigma_1  e^{ \, i b \sigma_3} 
	\qquad  b\in [0, 2\pi),
\end{equation}
which cannot be  continuously connected with the identity.
The general solution of the first type (\ref{one}) is
\begin{equation}\nonumber
\varphi'= -i\cot\frac{a}{2} \dot{\varphi},
\end{equation}
whereas 
\begin{align} \nonumber
&\operatorname{Re}(\dot\varphi +\varphi')+ \operatorname{Re}(\dot\varphi -\varphi') \cos b=\operatorname{Im}(\dot\varphi -\varphi')  \sin b\\  \nonumber
&\operatorname{Im}(\dot\varphi +\varphi')- \operatorname{Im}(\dot\varphi -\varphi') \cos b=\operatorname{Re}(\dot\varphi -\varphi')  \sin b
\end{align}
is the general solution of the second type (\ref{two}).

Let us consider some particular cases of physical interest. 
\begin{itemize} 
\item[i)] $U_D=\mathbb{I}$: {\bf Static  boundary conditions}. This is a boundary condition of the first type with $a=0$ 
\begin{equation}\nonumber
\dot\varphi=0.
\end{equation}
\item[ii)] $U_N=-\mathbb{I}$: {\bf Neumann boundary conditions}. This is a boundary condition of the first type with $a=\pi$ 
\begin{equation}\nonumber
\varphi'=0.
\end{equation}
\item[iii)] $U_{\mathrm{c}}=\pm i \sigma_3$: {\bf Chiral boundary conditions}. These are boundary conditions of the first type with 
$a=\pm\frac{\pi}{2}$ 
\begin{equation}\nonumber
		\varphi'=\mp i\dot\varphi. 
		\end{equation}
\item[iv)] $U_{\mathrm{t}}=\pm\sigma_1$: {\bf Twisted boundary conditions}. These are  boundary conditions of the second type with 
$b=0$ or  $b={\pi}$ 
\begin{align}\nonumber
		\operatorname{Im} \varphi'&=0 \qquad \operatorname{Re}\dot  \varphi=0\\ \nonumber
		\operatorname{Re} \varphi'&=0 \qquad \operatorname{Im}\dot \varphi=0.
\end{align}
\item[iv)] $U_{\mathrm{tc}}=\pm\sigma_2$: {\bf Twisted  chiral boundary conditions}. These are  boundary conditions of the second type with 
$b=\frac{\pi}{2}$ or  $b=\frac{3\pi}{2}$ 
\begin{equation}\nonumber
		\operatorname{Re} \varphi'=\mp \operatorname{Im}  \varphi' \qquad  \operatorname{Im} \dot \varphi=\pm \operatorname{Re}\dot \varphi .\\
\end{equation}
\end{itemize}

\subsection{Charged scalar fields two parallel plates}

Let us consider  a complex scalar field confined between two parallel plates
$\Omega=\{(x^1,x^2,x^3)\Big| -L\leq x^3\leq L  \}$.
In this case homogeneous local boundary conditions which preserve energy must satisfy that
\begin{equation}\label{2bc}
	\varphi_1^{*\prime}\dot\varphi_1+\dot{\varphi_1^*}\varphi_1'-\varphi_2^{*\prime}\dot\varphi_2-\dot{\varphi_2^*}\varphi_2'=0,
\end{equation}
where $\varphi_1(x^1,x^2)=\phi(x^1,x^2,-L)$ and $\varphi_2(x^1,x^2)=\phi(x^1,x^2,L)$ .
Written in terms of  the following vectors
\begin{equation}
	H_1=\begin{pmatrix} \nonumber
		\dot \varphi_1+\varphi_1'\\
		\dot \varphi_2-\varphi_2'\\
		\dot {\varphi_1^*}+\varphi_1^{*\prime}\\
		\dot {\varphi_2^*}-\varphi_2^{*\prime}
	\end{pmatrix} \hspace{0.5cm} H_2=\begin{pmatrix}
		\dot \varphi_1-\varphi_1'\\
		\dot \varphi_2+\varphi_2'\\
		\dot {\varphi_1^*}-\varphi_1^{*\prime}\\
		\dot {\varphi_2^*}+\varphi_2^{*\prime}
	\end{pmatrix}
\end{equation}
the restriction (\ref{2bc}) reads
\begin{equation}\nonumber
	\abs{H_1}^2-\abs{H_2}^2=4\left(\dot{\varphi^*_1}\varphi'_1+\varphi_1^{*\prime}\dot\varphi_1-\dot{\varphi^*_2}\varphi'_2-\varphi_2^{*\prime}\dot\varphi_2\right)=0.
\end{equation}
This means that any solution has to be of the form
\begin{equation}\label{eq_2_pared}
	H_1=UH_2
\end{equation}
with $U$ an unitary matrix of $U(4)$. There is another requirement  that this solution 
must satisfy.  Indeed, if we conjugate (\ref{eq_2_pared}) we get 
\begin{equation}\nonumber
	H_1^*=U^*H_2^*
\end{equation}
that implies  
\begin{equation}\nonumber
	 H_1=\begin{pmatrix}
		0&{\mathbb I}_2\\
		{\mathbb I}_2&0 
	\end{pmatrix}
	U^*\begin{pmatrix}
		0&{\mathbb I}_2\\
		{\mathbb I}_2&0 
	\end{pmatrix}H_2,
	\end{equation} 
which implies a further restriction on the unitary matrix 
\begin{equation}\label{conjugate_con}
	U=\begin{pmatrix}
		0&{\mathbb I}_2\\
		{\mathbb I}_2&0 
	\end{pmatrix} U^*\begin{pmatrix}
		0&{\mathbb I}_2\\
		{\mathbb I}_2&0 
	\end{pmatrix}.
\end{equation}
The meaning of this restriction is that the unitary matrix $U$ has to belong also to the $O(2,2)$ rotation group, because 
from (\ref{conjugate_con}) it follows that  $U$ leaves the $(2,2)$ metric 
\begin{equation}\nonumber
\begin{pmatrix}
		0&{\mathbb I}_2\\
		{\mathbb I}_2&0 
	\end{pmatrix}
\end{equation}
invariant. i.e.
\begin{equation}\nonumber
U^\perp \begin{pmatrix}
		0&{\mathbb I}_2\\
		{\mathbb I}_2&0 
	\end{pmatrix}  U= \begin{pmatrix}
		0&{\mathbb I}_2\\
		{\mathbb I}_2&0 
	\end{pmatrix}.
\end{equation} 

Thus, the general solution of the homogeneous local boundary conditions  is  given by the six-dimensional subgroup 
of  unitary matrices $G=O(2,2)\cap U(4)$  which has two disjoint components 
$G=O(2,2)\cap U(4)=G_+\cup G_-$ distinguished  by the sign of the determinant $ \det U=\pm 1$, i.e. the   component $G_+$
contains all the matrices of the form
\begin{equation}\label{double_one} 
	U_+(\mathbf a, \mathbf  b,\mathbf  c)= {\mathrm {exp}}\,  i
	\begin{pmatrix} 
		a_1&b_1+ib_2& 0 & c_1 + i c_2\\
		b_1-ib_2&a_2&-c_1 - i c_2& 0\\
		0&-c_1 + i c_2&-a_1&- b_1+ib_2 \\
		c_1 - i c_2&0& -b_1-ib_2&- a_2
	\end{pmatrix}
\end{equation}
with $\mathbf a, \mathbf  b,\mathbf  c\in \mathbb{R}^6$, that are continuously connected with the identity. The other component is given by the matrices of the form
\begin{equation} \label{double_two}
	U_-(\mathbf a, \mathbf  b,\mathbf  c)=\frac12 \begin{pmatrix} 
		1&1& -1 & 1 \\
		1&1&1& -1\\
		-1&1 &1&1\\
		1&-1& 1&1
	\end{pmatrix}  U_+(\mathbf a, \mathbf  b,\mathbf  c)
\end{equation}
that are disconnected from $G_+$. In summary, the homotopy group of $G$ is  $\pi_0(G)={\mathbb{Z}}_2$.

This group contains the solutions of the type considered in the previous case for each of the plane boundaries.
But the fact that there are two boundaries gives rise to other  remarkable boundary conditions like periodic boundary conditions.

Particular cases of interest are:
\begin{itemize}
\item[i)] $U_N=  {\mathbb I}_4$: {\bf Neumann boundary conditions}. 
i.e.  Neumann boundary conditions for both walls  $\varphi_1'=\varphi_2'=0$.
\item[ii)] $U_D=-{\mathbb I}_4$: {\bf Static boundary conditions}.  
i.e. static boundary conditions for both walls  $\dot\varphi_1=\dot\varphi_2=0$.

\item[iii)] 
$U_p=\begin{pmatrix} \sigma_1&0\\
0&\sigma_1
\end{pmatrix}:
$ {\bf Periodic boundary conditions}
 connecting the two walls
 $\dot\varphi_1=\dot\varphi_2$,  $\varphi_1'=\varphi_2'$, and

\item[iv)] 
$U_{\mathrm {ap}}=\begin{pmatrix} -\sigma_1&0\\
0&-\sigma_1
\end{pmatrix}$:  {\bf Antiperiodic boundary conditions}
 connecting the two walls in a different manner
 $\dot \varphi_1=-\dot \varphi_2$,  $\varphi_1'=-\varphi_2'$.

\end{itemize}

\section{Compatibility between energy and  charge conservations}
In the case of complex fields there is another conserved quantity, the electric charge. The charge conservation law provides another condition to be preserved by boundary conditions \cite{Bangalore}. In principle the families of boundary conditions which preserve charge and energy are different. However, in the case of charged scalar fields both families of boundary conditions are compatible. This very relevant property is a consequence of the compatibility of gauge transformations  and space-time traslations.  Indeed,  the actions of the $U(1)$  group of gauge transformations
\begin{equation}
G(\alpha)\phi= e^{i\alpha }\phi
\end{equation}
and the group of translations $T_4$
\begin{equation}
T_4 \phi(x)= \phi(x-a)
\end{equation}
 do commute.  
Moreover, a consequence of that property is that the Poisson bracket of the charge density \footnote{ $\Pi$ and $\Pi^\ast$ are  the canonical momenta
\begin{equation}
\Pi=\frac{\partial \mathcal L}{\partial \dot\phi}=\dot{\phi^*}\hspace{0.5cm}\Pi^*=\frac{\partial \mathcal L}{\partial \dot{\phi^*}}=\dot\phi
\end{equation}}
\begin{equation}\nonumber
\rho=\frac{i}{2}\left(\phi^*\dot\phi-\phi\dot{\phi^*}\right)=\frac{i}{2}\left(\phi^*\Pi^*-\Pi\phi\right)
\end{equation}
and the energy density
\begin{align}\nonumber
{\mathcal E}&=\frac{1}{2}\left(\dot{\phi^*}\dot{\phi}+\nabla \phi^*\nabla\phi\right)+ \frac{1}{2} m^2 \phi^2 + V(\phi)\\
&=\frac{1}{2}\left(\Pi\Pi^*+\nabla \phi^*\nabla\phi\right) + \frac{1}{2} m^2 \phi^2 + V(\phi)
\end{align}
vanishes, i.e. 
\begin{align}\nonumber
\{\rho,{\mathcal E}\}&=0.
\end{align}
The conservation of electric charge is given by the continuity equation
\begin{equation}\nonumber
\partial_t \rho + \partial^i j_i=0,
\end{equation}
where 
\begin{equation}\nonumber
{\mathbf j}=\frac{i}{2}\left(\phi^*{\boldsymbol \nabla} \phi-({\boldsymbol \nabla} \phi^*)\phi\right).
\end{equation}
In other words, charge  conservation requires the vanishing of the electric current flux through the boundary \cite{Bangalore}
\begin{equation}\nonumber
-\int_\Omega \dot \rho\ d^3 x=\int_\Omega\partial^i j_i\ d^3 x= \int_{\partial\Omega}{\mathbf j}\, d{\boldsymbol  \sigma}=\frac{i}{2}\int_{\partial \Omega}\left(\varphi^*{\boldsymbol \nabla} \varphi-({\boldsymbol \nabla} \varphi^*)\varphi\right)d{\boldsymbol \sigma}
\end{equation}
which in the right half space case reduces to
\begin{equation}\nonumber
\int_\Omega \dot \rho\ d^3 x=\frac{i}{2}\iint\left(\varphi^*\varphi' -\varphi^{*\prime}\varphi\right)dx^1 dx^2.
\end{equation}\label{chargecon}
Thus, homogeneous local  boundary conditions must satisfy
\begin{equation} \label{bcch}
\varphi^*\varphi' -\varphi^{\ast}{}'\varphi=0,
\end{equation}
and the most general boundary condition that satisfies  this constraint (\ref{bcch}) is given by \cite{aim} 
\begin{equation}\label{bcc}
\varphi+i\varphi' = U_c(\varphi - i \varphi'),
\end{equation}
in terms of an arbitrary unitary matrix $U_c$ of $L^2(\mathbb{R}^2)$\footnote{There is a technical subtlity associated to the fact that in higher dimension the baundary values $\varphi$ of the fields $\phi$ are singular which can be solved with a slight modification of the theory  \cite{amc} }.
Now if $\phi$ is a monochromatic field
\begin{equation} \nonumber
\phi= e^{i\omega t} \chi(x^1,x^2),
\end{equation}
 we have that
\begin{equation}\nonumber
\dot\phi=i\omega \phi\hspace{0.5cm} \dot{\phi^*}=-i\omega \phi^*.
\end{equation}
Thus, for monochromatic fields the vanishing condition associated to the conservation of energy (\ref{energycon}) implies the  conservation of charge (\ref{chargecon})
\begin{equation} \label{seis} \nonumber
\dot{\varphi^*}\varphi'+\varphi^{*\prime}\dot\varphi=-i\omega\left(\varphi^*\varphi'-\varphi\varphi^{*\prime}\right)=0,
\end{equation} 
and viceversa, the conservation of charge implies the conservation of energy.
 
However, the fact that   $U$ is independent of $\omega$   
does not guarantee that  boundary conditions 
which preserve the energy also preserve the electric charge  for multifrequency fields (\ref{bcc}).
 Only  boundary conditions  of the type
 \begin{equation} \nonumber
	U=\begin{pmatrix}
		U_c&0\\
		0&U_c^\ast
		\end{pmatrix},
	\label{blockunitary}
\end{equation}
 where $U_c$
has all eigenvalues $\pm 1$  define boundary conditions that preserve energy and  electric charge simultaneously. 
This type of boundary conditions includes Dirichlet, Neumann or periodic boundary conditions \cite{adj2} .

The application to gauge fields  is straightforward as we shall see below. However, the  methods based on energy consevation fail when applied to gravitation and  other higher spin gauge fields fields \cite{aadga}.

\section{Yang-Mills theory } \label{dkd}
Let us now consider  the interesting and non-trivial case of Yang-Mills gauge theories. 
The Lagrangian density  is given by 
\begin{equation}
	\mathcal{L}=-\frac{1}{16\pi}\text{tr } F_{\mu\nu}F^{\mu\nu}
\end{equation}
where $F_{\mu\nu}=\partial_\mu A_\nu-\partial_\nu A_\mu-[A_\mu,A_\nu]$,  $A_\mu$ being the gauge fields $A_\mu=A_\mu^aT^a$ and $T^a$ the generators of the Lie algebra. The symmetric energy-momentum tensor has the form
\begin{equation}
	T_{\mu\nu}=\frac{1}{4\pi}\text{tr }\left(\frac12\eta_{\nu\beta}F_{\mu\alpha}F^{\alpha\beta}+\frac12\eta_{\mu\beta}F_{\nu\alpha}F^{\alpha\beta}+\frac{1}{4}\eta_{\mu\nu}F_{\alpha\beta}F^{\alpha\beta}\right).
\end{equation}
Applying the translation symmetry as we did for the scalar case we have the following conservation energy law 
\begin{equation}\label{energy_law_ym}
	\partial^{\mu}T_{\mu 0}=\partial_t \mathcal E+\partial^i P_i=0,
\end{equation}
for the energy density 
\begin{equation} \nonumber
	\mathcal E=\frac{1}{8\pi}\text{tr }(\textbf E^2+\textbf B^2)
\end{equation} and the non-abelian Poynting vector
\begin{equation} \nonumber
	\textbf P=\frac{1}{4\pi}\text{tr } \left(\textbf E\times  \textbf B\right),
\end{equation}
where $E_i=F_{0i}$ and $B_i=-\frac{1}{2}\epsilon_{ijk}F_{jk}$ are  the chromo-electric and chromo-magnetic fields for the non-abelian case. Thus,  for any bounded domain $\Omega$ with regular boundary $\partial\Omega$ we have
\begin{equation}\nonumber
	\frac{d}{dt}E_\Omega=\int_\Omega \partial_t\mathcal E\ d^3x=-\int_\Omega \partial_i P^i\ d^3x=-\int_{\partial \Omega}n_i P^i\ d \sigma_{_{\partial \Omega}},
\end{equation}
where $\textbf n=n^i$ is the normal vector to the boundary surface $\partial \Omega$.

\subsection{Gauge fields in  half space}

As in the scalar theory, we consider the half-space
 $\Omega=\{\textbf x=(x^1,x^2,x^3)| x^3\geq 0\}$ that has a boundary $\partial\Omega=\{\textbf x=(x^1,x^2,0)\}$. The conservation of energy is given in this case by
\begin{equation}\nonumber
	\frac{d}{dt}E_\Omega=\iint T_{30}\  dx^1dx^2=\iint \text{tr }(B_2E_1-B_1E_2)dx^1dx^2,
\end{equation}
Restricting ourselves to  homogeneous local boundary conditions that are invariant under translation along $\partial\Omega$, the requirement  of energy conservation leads to 
\begin{equation}\label{cond_one_wall_ym}
	\text{tr }(B_2E_1-B_1E_2)=0.
\end{equation}
In terms of the following auxiliary vectors 
\begin{equation}\label{aux}
	H=\begin{pmatrix}
		E_1+iE_2+i(B_1+iB_2)\\
		E_1-iE_2-i(B_1-iB_2)
	\end{pmatrix}\hspace{.3cm} G=\begin{pmatrix}
		E_1-iE_2+i(B_1-iB_2)\\
		E_1+iE_2-i(B_1+iB_2)
	\end{pmatrix},
\end{equation}
the condition \eqref{cond_one_wall_ym} reads 
\begin{equation}\nonumber
	\text{tr }	(\abs{H}^2-\abs{G}^2)=-8\text{tr }(E_1B_2-E_2B_1),
\end{equation}
where we use that the trace of the product is conmutative. 
Thus, for homogeneous local boundary conditions the most general solution satisfying that
\begin{equation}\nonumber
	\abs{H}^2=\abs{G}^2
\end{equation}
 is 
\begin{equation}\label{sol_1_wall_ym_pre}
	H=U_b\otimes U_c\ G,
\end{equation}
where $U_b$ is an unitary 2x2 matrix  acting on the ${\mathbb C}^2$ chromoelectromagnetic vector fields $G$ and $U_c$ is an unitary matrix that acts on the color component of the fields. Since $F_{\mu\nu}$ is covariant under gauge transformations $\Phi$, $F_{\mu\nu}\rightarrow \Phi^{-1}F_{\mu\nu}\Phi$, $E$ and $B$ are always  gauge covariant, but not gauge invariant 
 and therefore the most general solution of  \eqref{sol_1_wall_ym_pre} 
is not gauge invariant
 \begin{equation}\label{gauge_sol}
 \Phi^{-1}H\Phi=\Phi^{-1} U_b\otimes U_c\ G\Phi\neq U_b\otimes U_c\ \Phi^{-1}G\Phi.
\end{equation}
 However, all physically consistent boundary conditions must be gauge invariant. This implies that the only physically possible boundary conditions are those where 
 $U_c$ is an element of the center of the gauge group, i.e.  $U_c\in Z(G)$. In such a case the 
 transformation \eqref{gauge_sol} can be rewritten as
\begin{equation}\nonumber
 \Phi^{-1}H \Phi=U_b\otimes U_c\ \Phi^{-1} G \Phi
\end{equation}
and  the corresponding boundary conditions of gauge fields 
are gauge invariant. In the case where the gauge group is $SU(N)$ we have that the center of the group has the form
\begin{equation}\nonumber
Z(SU(N))=\{e^{\frac{2\pi i n}{N}}I_N; n=0,1\ldots,N-1\}.	
\end{equation} Using this expression we can write the boundary condition \eqref{sol_1_wall_ym_pre} as
\begin{equation}\label{sol_1_wall_inv}
	H=e^{\frac{2\pi i n}{N}}U_b\otimes I_N\ G.
\end{equation}
This extra phase that the color adds can be reabsorbed into $U_b$ since it does not affect the condition of being unitary. Thus, we can rewrite \eqref{sol_1_wall_inv} as
\begin{equation}\label{sol_1_wall_ym} \nonumber
	H=U\otimes I_N G,
\end{equation}
where $U=e^{\frac{2\pi i n}{N}}U_b $ is an unitary 2x2 matrix. If we conjugate this relation we have 
\begin{equation}\nonumber
	\sigma_1H=U^*\otimes I_N\sigma_1 G,
\end{equation}
and we get that the unitary 2x2 matrix has to satisfy  the extra condition
\begin{equation}\nonumber
	U=\sigma_1U^*\sigma_1.
\end{equation}
Thus, we get the same type of matrices as for the charged scalar field in half space with the two disjoint components \eqref{one} and \eqref{two}. In the first case \eqref{one} the local boundary conditions  are given by
\begin{equation}
\label{solp_one_wall_ym}
	\begin{pmatrix}
		E_2\\
		B_2
	\end{pmatrix}=\tan\frac{a}{2} 	\begin{pmatrix}
		E_1\\
		B_1
	\end{pmatrix},
\end{equation}
whereas in the second case \eqref{two} they are
\begin{equation}\label{soln_one_wall_ym}
	\begin{pmatrix}
		B_1\\
		B_2
	\end{pmatrix}=-\tan\frac{b}{2}\begin{pmatrix}
		E_1\\
		E_2
	\end{pmatrix}.
\end{equation}
Some interesting particular cases  are
\begin{enumerate}[i)]
	\item [i)] $U_+(0)=\mathbb{I}$ $\Rightarrow$ $E_2=B_2=0$.
	\item [ii)]  $U_+(\pi)=-\mathbb{I}$ $\Rightarrow$ $E_1=B_1=0$.
	\item $U_+(\pm \frac{\pi}{2})=\pm i\sigma_3$  $\Rightarrow$ $E_2=\pm E_1$ and $B_2=\pm B_1$.
	\item $U_-(0)=\sigma_1$ $\Rightarrow$ $B_1=B_2=0$.
	\item $U_-(\pi)=-\sigma_1$ $\Rightarrow$ $E_1=E_2=0$.
	\item $U_-(\pm\frac{\pi}{2})=\pm \sigma_2$ $\Rightarrow$  $B_1=\mp E_1$ and $B_2=\mp E_2$.
\end{enumerate}
\subsection{Boundary conditions for the Yang-Mills potentials}
The above boundary conditions  can be formulated in terms of  chromo-electromagnetic potentials. 


In the case  $U_-$ \eqref{soln_one_wall_ym} we have  the boundary conditions 
\begin{align}\label{gaugebc_1}
	&\partial_2 A_3-A'_2-[A_2,A_3]=-\tan \frac{b}{2} (\dot A_1-\partial_1 A_0-[A_0,A_1]),\\ \label{gaugebc_2}
	&\partial_1A_3- A'_1-[A_1,A_3]=\tan\frac{b}{2}(\dot A_2-\partial_2 A_0-[A_0,A_2]),
\end{align}
 in terms of $A_\mu$, where 
 \begin{equation}
	\dot A=\partial_t A|_{\partial \Omega}
\end{equation}
and 
\begin{equation}
	A'=\partial_3 A|_{\partial \Omega}. \label{gaugebc}
\end{equation}
The above boundary conditions \eqref{gaugebc_1} and \eqref{gaugebc_2} can also be rewritten using the covariant derivative $D_\mu A_\nu= \partial_\mu A_\nu-[A_\mu,A_\nu]$, 
\begin{align}\nonumber
	&D_2A_3-A_2'=\tan\frac{b}{2}\left(D_0A_1-D_1A_0-[A_0,A_1]\right),\\\nonumber
	&D_1A_3-A_1'=-\tan\frac{b}{2}\left(D_0A_2-D_2A_0-[A_0,A_2]\right).
\end{align}
Choosing as gauge fixing condition
\begin{equation}\nonumber
	A_0=0,\hspace{0.5cm}\partial_1 A_1+ \partial_2 A_2+A'_3=0,
\end{equation}
we can rewrite the conditions as
\begin{equation} \nonumber
	\begin{pmatrix}
		A_1'\\
		A_2'\\
		A_3'
	\end{pmatrix}=M \begin{pmatrix}
		A_1\\
		A_2\\
		A_3
	\end{pmatrix}, \label{cayley}
\end{equation}
where 
\begin{equation} M=
\begin{pmatrix}
		0& \tan\frac{b}{2}D_0& D_1\\
		-\tan\frac{b }{2}D_0&0& D_2\\
		-D_1& -D_2&0
	\end{pmatrix}
	\end{equation}
is a  selfadjoint differential operator that is gauge covariant and relates the normal derivatives of the gauge fields with  the very gauge fields in that gauge. The expression  \eqref{cayley} is reminiscent of the one obtained for scalar fields where the matrix $M$ is related to the family of unitary matrices by means of a Cayley transform \cite{aim}.

 In the case of the family of solutions \eqref{solp_one_wall_ym} given by $U_+$ we have 
\begin{align}\nonumber
	&\dot A_2-\partial_2 A_0-[A_0,A_2]=\tan\frac{a}{2}(\dot A_1-\partial_1 A_0-[A_0,A_1]),\\ \nonumber
	& \partial_2 A_3-A'_2-[A_2,A_3]=-\tan\frac{a}{2}(\partial_1A_3- A'_1-[A_1,A_3]),
\end{align}
that can be rewritten  using the covariant derivatives
\begin{align}\nonumber
	&D_0A_2-\partial_2A_0=\tan\frac{a}{2}(D_0A_1-\partial_1A_0)\\ \nonumber
	&\partial_2A_3-D_3A_2=-\tan\frac{a}{2}(D_1A_3-A'_1).
\end{align}
Choosing a diferent gauge fixing condition
\begin{equation}\nonumber
	A_1=0,\hspace{0.5cm}\partial_tA_0+\partial_2 A_2+ \partial_3A_3=0,
\end{equation}
these boundary conditions reduce to 
\begin{equation}\nonumber
	\partial_2\begin{pmatrix}
		A_0\\
		A_2\\
		A_3
	\end{pmatrix}=\begin{pmatrix}
		\tan\frac{a}{2}D_1&D_0&0\\
		-D_0& 0&-D_3&\\
		0& D_3&-\tan\frac{a}{2}D_1
	\end{pmatrix}\begin{pmatrix}
		A_0\\
		A_2\\
		A_3
	\end{pmatrix}
\end{equation}
in terms of  the selfadjoint differential operator that is gauge  covariant.

\section{Interacting theories of matter and gauge fields}

In previous sections we have  considered  scalar and gauge field theories separately.  But in the Standard Model they appear
interacting one with each other. The analysis of boundary conditions that preserve energy in that case 
can be carried out along the same lines.

Let us consider the case of scalar fields interacting with $SU(N)$ gauge fields
\begin{equation}
	\mathcal{L}=\frac{1}{2}(D_\mu\phi, D^\mu\phi) - \frac{1}{2} m^2 ||\phi||^2 - V( ||\phi||^2 )  -\frac{1}{16\pi}\text{tr } F_{\mu\nu}F^{\mu\nu} ,
\end{equation}
where $\phi$ is a scalar field supporting an irreducible unitary n-dimensional representation $\rho_n$  of  $SU(N)$, $D_\mu\phi=\partial_\mu \phi+ \widetilde{\rho_n}(A)\phi$   its covariant derivative,  $(,)$ denotes the  product of the scalar fields  associated to the
 unitary representation $\rho_n$, and
$V$ is any arbitrary local potential function.

The  current associated to the conservation of energy is the linear momentum whose components are
\begin{equation}
	T_{0 i}=\frac{1}{2}(D_0\phi, D_i\phi)+\frac{1}{2}(D_i\phi, D_0\phi)+\  \frac1{4 \pi} {\rm tr} E^j F_{ji}.
\end{equation}
The vanishing of the total linear momentum flux  across the boundary of the field domain guarantees  the conservation of the energy inside such a domain. In the case of a half space with an homogeneous boundary there are a large number of boundary conditions which satisfy this requirement, but we will only consider those where the flux of the two independent components of the current vanish separately,  i.e. the fluxes of  the gauge fields  and the $\phi$ fields    both vanish.
In such a case the boundary conditions are given by 

\begin{equation}\label{sol_1_wall_ym_preb}
	H=U_b\otimes U_c\ G,
\end{equation}
where $H$ and $G$ are  the gauge covariant generalization of the auxiliary fields defined by  (\ref{aux})
and 
\begin{equation} \label{bce1b}
	\begin{pmatrix}
		D_3\varphi+D_0\varphi\\
		D_3\varphi^{*}+D_0{\varphi^*}
	\end{pmatrix}=\widetilde{U}_b\otimes \widetilde{U}_c\  \begin{pmatrix}
		D_3\varphi-D_0\varphi\\
		D_3\varphi^{*}-D_0{\varphi^*}
	\end{pmatrix},
\end{equation}
where $ U_b\in U(2)$ and $\widetilde{U}_b\in U(2)$ are two $2\times 2$ unitary matrices and 
$ {U}_c\in U(N)$, $\widetilde{U}_c\in U(n)$ are two  unitary matrices associated  to the gauge group
representations involved in the theory.
As we have shown in section \ref{dkd} gauge covariance requires that  the last two matrices 
must belong to the center of their unitary groups, i.e.  ${U}_c\in {\mathbb Z}_N, \widetilde{U}_c\in {\mathbb Z}_n$
because of the irreducible character of the gauge group representations of  matter fields.
The corresponding phases can be absorbed into the matrices   $U_b, \widetilde{U_b}\in U(2)$. One might think  that some differences  should appear when one considers fields  either in 
the fundamental   or  the   adjoint  representation,
 but there is none. The reason being that both theories support irreducible representations of $SU(N)$. 


\section{Maxwell theory}
The analysis of previous sections include Maxwell theory of electromagnetic fields since  it is a $U(1)$ gauge theories . In this case the chromoelectromagentic fields become the standard electromagnetic fields, and  covariant derivatives are replaced by standard derivatives in the  selfadjoint differential operators that appear in the motion equations. 

An interesting case of study  is the analysis of the boundary conditions on the interface between two different electromagnetic active media. Let us consider a linear, non dispersive and isotropic case where $D=\epsilon E$ and $B=\mu H$, with $\epsilon$ the dielectric permitivity and $\mu$ the magnetic permeability of the material, and work on Gaussian units. In this type of media the electromagnetic energy density is given by
\begin{equation}
\mathcal E=\frac{1}{8\pi}(\textbf E\cdot \textbf D +\textbf B\cdot \textbf H)
\end{equation}
and the Poynting vector by
\begin{equation}
\mathbf S=\frac{c}{4\pi}\mathbf E \times \mathbf H,
\end{equation}
where we recovered the dependency on the vacuum speed of light $c$. The energy conservation law \eqref{energy_law_ym} becomes
\begin{equation}
\partial_t\mathcal E +\nabla \textbf S=0 .
\end{equation}

If we consider the case of two different electromagnetic active media in  $\Omega_+=\{\textbf x=(x^1,x^2,x^3)| x^3\geq 0\}$ and $\Omega_-=\{\textbf x=(x^1,x^2,x^3)| x^3\leq 0\}$ which are separated by the boundary $\partial\Omega=\{\textbf x=(x^1,x^2,0)\}$, the condition for conservation of energy for both media is given by
\begin{equation} \nonumber
	\frac{d}{dt}(E_{\Omega_+}+E_{\Omega_-})=\int_{\Omega_+} \partial_t\mathcal E\ d^3x\ +\int_{\Omega_-} \partial_t\mathcal E\ d^3x=\int_{\partial\Omega_{+}}S^3 dx^1dx^2\ -\int_{\partial\Omega_{-}}S^3 dx^1dx^2,
\end{equation}
where $\partial\Omega_+$ is the boundary at the right side and $\partial\Omega_-$ is the boundary at the left side. Inserting the explicit expression of the Poynting vector for both materials the condition has the form
\begin{equation}\nonumber
	\frac{c}{4\pi\mu^+}\int_{\partial {\Omega_+}}\!\!\! (E_1^{+}B_2^{+}-E_2^{+}B_1^{+})\ dx^1dx^2-\frac{c}{4\pi\mu^-}\int_{\partial \Omega_-}\!\! \!(E_1^{-}B_2^{-}-E_2^{-}B_1^{-})\ dx^1dx^2=0
\end{equation}
where  $\mu^\pm$ is the magnetic permeability in each media, and $E^\pm$ and $B^\pm$ are the electromagnetics field on each side. Considering only homogeneous boundary conditions that are invariant under translations along the boundary  plane  $\partial\Omega$, the energy conservation condition reduces to
\begin{equation}
\frac{1}{\mu^+}(E_1^{+}B_2^{+}-E_2^{+}B_1^{+})-\frac{1}{\mu^-}(E_1^{-}B_2^{-}-E_2^{-}B_1^{-})=0.
\end{equation} 
We can define two auxiliary vectors
\begin{align}\nonumber
	&H=\begin{pmatrix}
		\frac{1}{\sqrt{\mu^-}}\left(E_1^- +iE_2^-+i(B_1^-+iB_2^-)\right)\\
		\frac{1}{\sqrt{\mu^+}}\left(E_1^+-iE_2^++i(B_1^+-iB_2^+)\right)\\
		\frac{1}{\sqrt{\mu^-}}\left(E_1^--iE_2^- -i(B_1^- -iB_2^-)\right)\\
		\frac{1}{\sqrt{\mu^+}}\left(E_1^++iE_2^+-i(B_1^+ +iB_2^+)\right)
	\end{pmatrix},\\\nonumber
		& G=\begin{pmatrix}
		\frac{1}{\sqrt{\mu^-}}\left(E_1^- -iE_2^- +i(B_1^- -iB_2^-)\right)\\
		\frac{1}{\sqrt{\mu^+}}\left(E_1^+ +iE_2^+ +i(B_1^+ +iB_2^+)\right)\\
		\frac{1}{\sqrt{\mu^-}}\left(E_1^-+iE_2^--i(B_1^-+iB_2^-)\right)\\
		\frac{1}{\sqrt{\mu^+}}\left(E_1^+-iE_2^+-i(B_1^+-iB_2^+)\right)
	\end{pmatrix},
\end{align} 
in terms of which the condition reads
\begin{equation} \nonumber
		\abs{H}^2-\abs{G}^2=
		\frac{8}{\mu^+}E_1^+B_2^+ -\frac{8}{\mu^+}E_2^+B_1^+-\frac{8}{\mu^-}E_1^-B_2^-+\frac{8}{\mu^-}E_2^-B_1^-.
\end{equation}
Thus  the general solution of the boundary condition is given by 
\begin{equation} \nonumber
H=UG,
\end{equation}
where $U$ is a 4x4 unitary matrix. Gauge invariance is automatic since the elements of $U(1)$ are simply a phase $e^{i\beta}$ which can be absorbed in the unitary matrix.  Conjugating this equation leads to
\begin{equation}\nonumber
	\begin{pmatrix}
		0&\mathbb{I}\\
		\mathbb{I}& 0
	\end{pmatrix}H=U^*\otimes I_N\begin{pmatrix}
		0&\mathbb{I}\\
		\mathbb{I}& 0
	\end{pmatrix}G,
\end{equation}
which gives an extra constraint for the unitary matrix
\begin{equation}\label{extracond_two_wall} \nonumber
	U=\begin{pmatrix}
		0&\mathbb{I}\\
		\mathbb{I}& 0
	\end{pmatrix}U^*\begin{pmatrix}
		0&\mathbb{I}\\
		\mathbb{I}& 0
	\end{pmatrix}.
\end{equation} 
Thus, we get the same 4x4 unitary  matrix as for the scalar charged fields for the two parallel plates, which are given by the two disjoint components \eqref{double_one} and \eqref{double_two}. Let see some examples of this boundary conditions:
\begin{enumerate}[i)]
	\item$U_+\left(0,0,0,\frac{\pi}{2},0,-\frac{\pi}{2}\right)=\begin{pmatrix}
		0&I\\
		I&0
	\end{pmatrix}$. In this case the boundary magnetic fields vanish in both sides $B_1^+=B^+_2=B_1^-=B_2^-=0$.
	\item$U_+\left(0,0,0,\frac{\pi}{2},0,\frac{\pi}{2}\right)=-\begin{pmatrix}
		0&I\\
		I&0
	\end{pmatrix}$. In this case the boundary electric fields vanish in both sides $E_1^+=E_2^+=E_1^-=E_2^-=0$.
	\item $U_+\left(\frac{\pi}{2},\frac{\pi}{2},\mp\frac{\pi}{2},0,0,0\right)=\pm\begin{pmatrix}
		\sigma_1&0\\
		0&\sigma_1
	\end{pmatrix}$. In this case $E_1^+=\pm\alpha E_1^-$, $E_2^+= \pm\alpha E_2^-$, $B_1^+= \pm \alpha B_1^-$ and $B_2^+= \pm \alpha B_1^-$, which leads to periodic/antiperiodic boundary conditions 
with a jump given by $\alpha=\sqrt{\frac{\mu^+}{\mu^-}}$.
\end{enumerate}
 These boundary conditions include those obtained in the case of one simple wall, like in the first and second case, where simply the Poynting vector is zero at each side of the boundary so there is no flux of energy through the materials. The new phenomena due to the two different materials occurs in the boundary conditions where the fluxes of energy from each side are not zero. The constraint of energy conservation induces in this case a jump in the values of the electromagnetic fields related to the different  magnetic permeabilities of the materials. This can be seen in the periodic and antiperiodic boundary conditions showed above, where instead of just getting the same values at the sides of the boundary there is a jump on them across the boundary.

\section{Conclusions}

We have developed a consistent theory of boundary conditions for relativistic field theories based on  the conservation of fundamental quantities: charge and energy. These are the two  generic conservation laws of Nature. Boundary conditions which preserve charge are quite well known \cite{aim}--\cite{Bangalore} whereas those that preserve the energy have not been explored so deeply. We have analyzed in this paper the theory of boundary conditions from this perspective.
We  found infinite families of energy preserving boundary conditions in scalar, electromagnetic and non-abelian gauge theories. In some cases the boundary conditions preserve both charge and energy. In the Maxwell  and  Yang-Mills theory
the boundary conditions are also gauge invariant. We also have shown how this method can be used to describe boundaries between different material media applying the conservation of energy.

Spinor fields have not been considered in this paper but the extension of the analysis to this case is straightforward. However, the generalization for gravitation and higher spin fields is not so simple. In the gravitational case the difficulty resides on the fact that the vanishing condition required to define the appropriate boundary conditions is automatically satisfied by any solution of classical equation of motion.
In this case a generalization of the application of the above  conservation laws is required. 

Another avenue worth to explore is the extension of this analysis for regularizations of field  theories on the lattice. This is very interesting for numerical approaches both for the classical dynamics of interacting field theories and for the analysis of  non-perturbative  effects in the corresponding quantum theories.

One of the interesting boundary effects associated to the boundary conditions  is the dependence of the non-perturbative Casimir energy in 2+1 gauge theories on the boundary conditions \cite{Chernodub}\cite{NK}. The exponential decay of the Casimir energy with the distance between two parallel plates  is dependent on the boundary conditions and points out to a new mass parameter much lower than the glueball mass, opening the door to new interpretations of the confinement mechanism.

\section*{Acknowledgments}
M. A. thanks to A.P. Balachandran, A. Ibort and  G. Marmo for so many  discussions.

We are  partially supported by Spanish  Grant PGC2022-126078NB-C21 funded by MCIN/AEI/ 10.13039/501100011033 and “ERDF A way of making EuropeGrant , the DGA-FSE grant 2020-E21-17R of the Aragon Government 
and the European Union - NextGenerationEU Recovery and Resilience Program on 'Astrofísica y Física de Altas Energías' CEFCA-CAPA-ITAINNOVA

\end{document}